\documentclass[12pt]{iopart}

\usepackage{graphicx}
\usepackage{epstopdf}
\usepackage{enumitem}

\usepackage{iopams}
\begin{document}

\title[]{Evidence of Cluster Structure of $^9$Be from $^3$He+$^9$Be Reaction}

\author{S~M~Lukyanov$^1$, M~N~Harakeh$^2$, M~A~Naumenko$^1$, Yi~Xu$^3$, W~H~Trzaska$^4$,  V~Burjan$^3$, V~Kroha$^3$, J~Mrazek$^3$, V~Glagolev$^3$, \v{S}~Pisko\v{r}$^3$, E~I Voskoboynik$^1$, S~V~Khlebnikov$^{5}$, Yu~E~Penionzhkevich$^{1,6}$, N~K~Skobelev$^1$, Yu~G~Sobolev$^1$, G~P~Tyurin$^3$, K~Kuterbekov$^{7}$ and Yu~Tuleushev$^8$}
\address{$^1$ Flerov Laboratory of Nuclear Reactions, Dubna, Russian Federation}
\address{$^2$ KVI-CART, University of Groningen, Groningen, Netherlands}
\address{$^3$ Nuclear Physics Institute, \v{R}e\v{z}, Czech Republic}
\address{$^4$ Department of Physics, University of Jyv\"askyl\"a, Jyv\"askyl\"a, Finland}
\address{$^5$ Khlopin Institute, St. Petersburg, Russian Federation}
\address{$^6$ National Research Nuclear University ``MEPhI'', Moscow, Russian Federation}
\address{$^7$ Eurasian Gumilev University, Astana, Kazakhstan}
\address{$^8$ Nuclear Physics Institute, Almaty, Kazakhstan}

\ead{lukyan@nrmail.jinr.ru}
\begin{abstract}
The study of inelastic scattering and multi-nucleon transfer reactions was performed by bombarding a $^{9}$Be target with a $^3$He beam at an incident energy of 30 MeV. Angular distributions for $^9$Be($^3$He,$^3$He)$^{9}$Be, $^9$Be($^3$He,$^4$He)$^{8}$Be, $^9$Be($^3$He,$^5$He)$^{7}$Be, $^9$Be($^3$He,$^6$Li)$^6$Li and $^9$Be($^3$He,$^5$Li)$^7$Li reaction channels were measured. Experimental angular distributions for the corresponding ground states (g.s.) were analysed within the framework of the optical model, the coupled-channel approach and the distorted-wave Born approximation. Cross sections for channels leading to unbound $^5$He$_{g.s.}$, $^5$Li$_{g.s.}$ and $^8$Be systems were obtained from singles measurements where the relationship between the energy and the scattering angle of the observed stable ejectile is constrained by two-body kinematics. Information on the cluster structure of $^{9}$Be was obtained from the transfer channels. It was concluded that cluster transfer is an important mechanism in the investigated nuclear reactions. In the present work an attempt was made to estimate the relative strengths of the interesting $^8$Be+$n$ and $^5$He+$\alpha$ cluster configurations in $^9$Be. The branching ratios have been determined confirming that the $^5$He+$\alpha$ configuration plays an important role. The configuration of $^9$Be consisting of two bound helium clusters $^3$He+$^6$He is significantly suppressed, whereas the two-body configurations ${}^{8}$Be+$n$ and ${}^{5}$He+$\alpha$ including unbound $^8$Be and $^5$He are found more probable.
\end{abstract}


\section{Introduction}

In recent years the study of light radioactive nuclei \cite{Ref1,Ref2} has intensified due to the significant progress made with radioactive beam facilities. It has led to a decrease of interest in the study of light stable nuclei such as ${}^{6,7}$Li and ${}^{9}$Be. It has been shown that in light nuclei the nucleons tend to group into clusters, the relative motion of which defines to a large extent the properties of these nuclei. Consequently, the cluster structure of their ground as well as low-lying excited states became the focus of theoretical and experimental studies. For example, ${}^{6}$Li and ${}^{7}$Li nuclei are both well described by the two-body cluster models ($\alpha+d$ and $\alpha+t$, respectively).

Due to its Borromean structure, a special attention has been focused on the ${}^{9}$Be nucleus which may be considered as a nuclear system with the following two-body configurations: ${}^{8}$Be+$n$ or ${}^{5}$He+$\alpha$.

The breakup of $^9$Be via $^8$Be$_{g.s.}$ has been measured for many of the low-lying excited states of $^9$Be \cite{Brown, Papka}. However, the breakup branching via the first-excited 2$^+$ state of $^8$Be and via $^5$He remained uncertain.

The three-body configuration ($\alpha\alpha n$) of $^9$Be may have a significant astrophysical interest. The short lifetimes of $^8$Be$_{g.s.}$, the first-excited state of $^8$Be and $^5$He$_{g.s.}$ suggest that the sequential capture of a neutron or an $\alpha$ particle is very unlikely. The formalism used to derive the ($\alpha\alpha n$) rate by Grigorenko~\cite{Grigorenko} also suggested that any broad intermediate resonances will have little effect on the ($\alpha\alpha n$) rate. However, according to another theoretical calculation \cite{Buchmann} the stellar reaction rate for this reaction proceeding via the $^5$He$_{g.s.}$ channel is significant for the formation of $^9$Be.

In the measurement \cite{Soic} of the $^9$Be($^7$Li,$^7$Li$\alpha$)$\alpha- n$ reaction it was unambiguously established for the first time that the $^9$Be excited states around 6.5 and 11.3 MeV decay into the $\alpha+^5$He channel.

In Ref.~\cite{Charity}, the structure of $^9$Be was discussed in the frame of isobaric analogue states of $^9$B. Charity et al. \cite{Charity} found that the decay of ${}^9$B has the dominant branch $\alpha$+${}^5$Li implying that ``the corresponding mirror state in ${}^{9}$Be would be expected to decay through the mirror channel $\alpha$+${}^{5}$He, instead of through the $n$+${}^{8}$Be(2$^+$) channel The mirror state of $^9$Be at 2.429 MeV is also reported to decay by $\alpha$ emission". In this case it would decay to the unstable ground state of $^5$He producing the $n$+2$\alpha$ final state. On the contrary, another experimental work~\cite{Papka} claimed that this state decays almost exclusively by $n$ emission to the unstable first-excited state of $^8$Be. 

The excited states of $^9$Be have been populated in various ways including among others $\beta$-decay \cite{Nyman}. Most experiments confirm that the 2.429 MeV state has a branching ratio to the $^8$Be$_{g.s.}$+$n$ channel of only $ \sim$7\% \cite{Brown}, but could not determine whether the remaining strength was in the $^8$Be($2^+$)+$n$ or the $^5$He+$\alpha$ channel. However, it was reported in Ref.~\cite{Nyman} that the ratio 2:1 could be assigned for the two channels, respectively.

The results of Refs.~\cite{Brown, Papka, Grigorenko, Soic, Charity} along with the qualitative breakup data discussed above suggest the necessity of obtaining quantitative branching-ratio data for the low-lying states in the Borromean nucleus $^9$Be.

The present experiment was designed to study the breakup of $^9$Be in the attempt to determine the contribution of $^5$He+$\alpha$ and $n$+$^8$Be channels. Our data are based on inclusive measurements, whereas the experimental results of Refs.~\cite{Brown, Papka} were obtained in exclusive measurements. The exclusive measurements require fully defined kinematics with complicated detector systems allowing the complete reconstruction and identification of the breakup event. The inclusive measurements, in spite of their simplicity, may also be useful in the study of clustering phenomena. For example, in our recent paper~\cite{luk} the obtained large value of the deformation parameter might be considered as the confirmation of the cluster structure of the low-lying states of $^9$Be. However, the inclusive measurement did not allow us to give unambiguous preference to one of the possible configurations, e.g. ($\alpha$+$\alpha$+$n$) or ($\alpha$+${}^{5}$He).

Another purpose of the present study was the attempt to find the cluster structure (for instance, $^5$He) and how the cluster structure influences the nuclear reaction mechanism.

Indeed, D\'{e}traz et al. \cite{Detraz, Detraz2} argued that multi-particle-multi-hole structures are expected to occur at rather low excitation energies in nuclei. Therefore, four-nucleon transfer reactions have been extensively studied. It is hoped that the major features of such reactions, in spite of the a priori complexity of such a transfer, may be understood assuming that the nucleons are transferred as a whole, i.e. strongly correlated in the cluster which has the internal quantum numbers of a free $\alpha$-particle.

Whether one or several internal configurations of the transferred fragments contribute to the transition amplitudes of these transfer reactions is both a crucial and difficult question to address. In the present study, we tried to examine how, for instance, two protons and two neutrons merged into an $\alpha$-cluster are transferred.

The angular distributions of the $^9$Be($^3$He,$^7$Be)$^5$He and $^9$Be($^3$He,$^6$Li)$^6$Li reaction channels were scrupulously measured at the energy of 60 MeV for the transitions to the ground states of $^5$He and $^6$Li by Rudchik et al.~\cite{Rudchik}. The experimental data were analysed using the coupled reaction channels (CRC) model including one- and two-step cluster transfer and cluster spectroscopic amplitudes calculated in the framework of the translation-invariant shell model. It was concluded that the cluster transfer is unimportant: in the $^9$Be($^3$He,$^7$Be)$^5$He reaction channel the $\alpha$-transfer dominates only at small angles while the transfer of two neutrons dominates at large angles. Nevertheless, the conclusion of Ref.~\cite{Rudchik} in general supports the idea of cluster transfer.

The second motivation for studying again the $^3$He+$^9$Be reaction was the attempt to find evidence for the cluster structure of these nuclear systems by the analysis of the reaction products. 

\section{Experimental procedure}

The $^3$He+$^9$Be experiments were performed at the K130 Cyclotron facility of the Accelerator Laboratory of the Physics Department of Jyv\"{a}skyl\"{a} University and at the Nuclear Physics Institute (NPI), \v{R}e\v{z}, Czech Republic. The $^3$He beam energy was 30 MeV. The average beam current during the experiments was maintained at 10 nA. The self-supporting Be target was prepared from a 99\% pure thin foil of beryllium. The target thickness was 12 $\mu$m. Peaks due to carbon and oxygen contaminations were not observed in the energy spectra.
\begin{figure}[htp]

  \centering

  \begin{tabular}{c}

    \includegraphics[width=90mm]{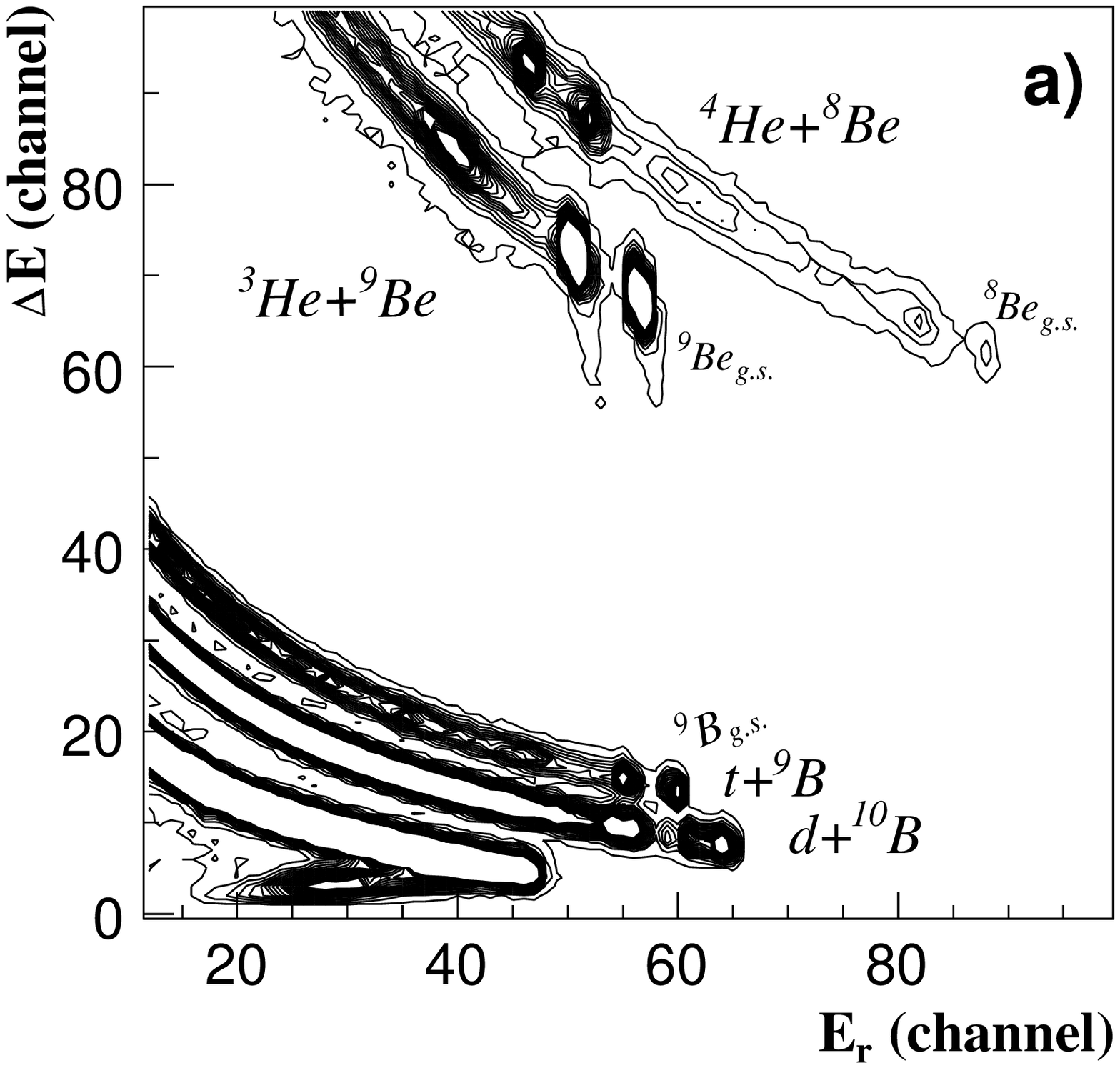}\\

    \includegraphics[width=85mm]{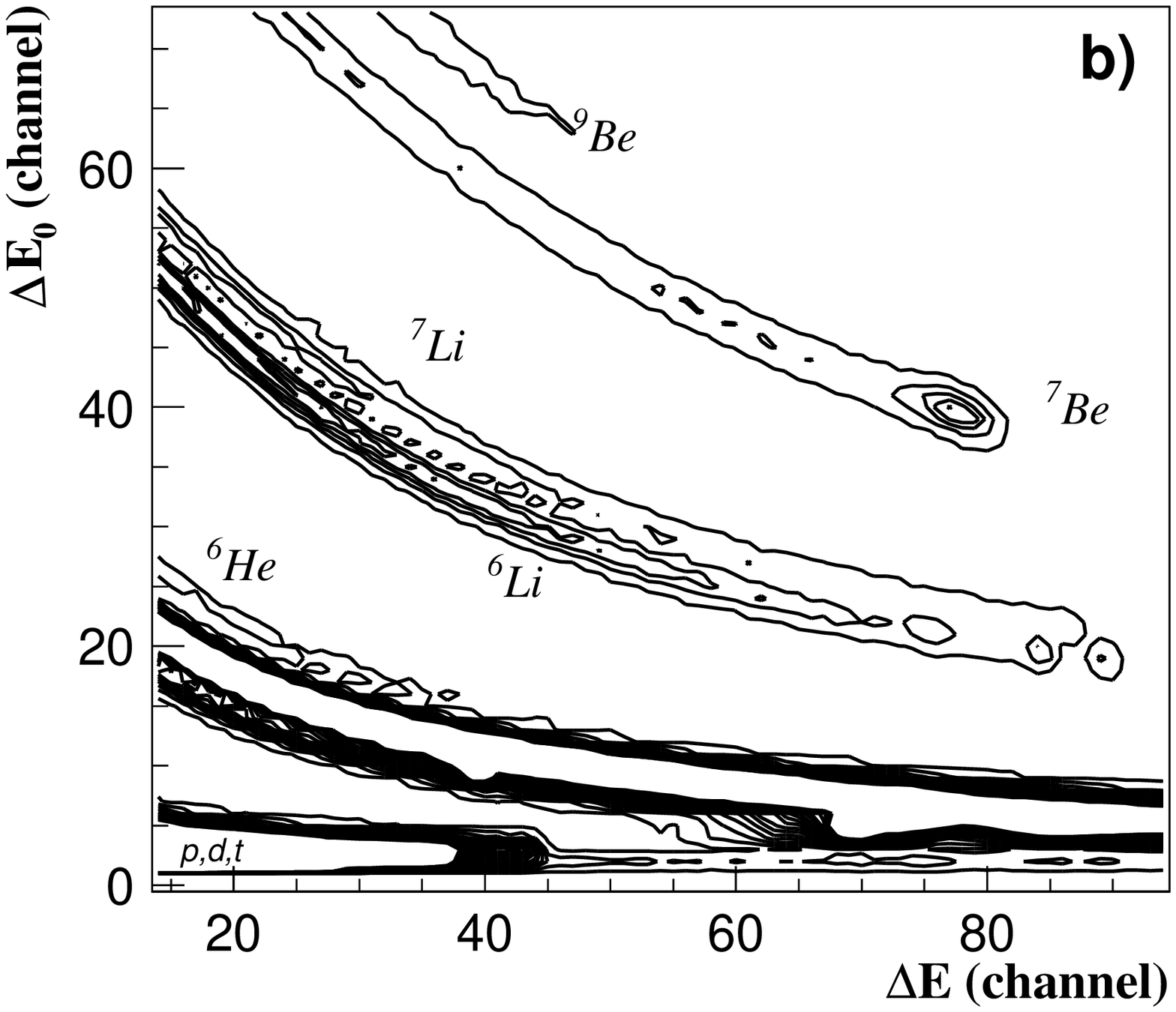}\\

  \end{tabular}

 \label{figure}\caption{
Particle identification plots for the products of the $^{3}$He+$^9$Be reaction: a) $p$, $d$, $t$ and $^{3,4}$He; b) $^{6}$He, $^{6,7}$Li and $^{7,9}$Be. $\Delta$E is the energy loss and E$_r$ is the residual energy. The loci for the products are indicated.
}\label{fig1}

\end{figure}

To measure (in)elastically scattered ions four Si-Si(Li) telescopes each consisting of $\Delta$E$_0$, $\Delta$E and E$_r$ detectors with thicknesses 10 $\mu$m, 100 $\mu$m and 3 mm, respectively, were used. Each telescope was mounted at a distance of $\sim$45 cm from the target. Particle identification was based on the measurements of energy-loss $\Delta$E and residual energy E$_r$ (the so-called $\Delta$E-E method). The telescopes were mounted on rotating supports, which allowed to obtain data from $\theta_{lab}$ = 20$^\circ$ to $\theta_{lab}$ = 107$^\circ$ in steps of 1-2$^\circ$. The overall energy resolution of the telescopes was $\sigma \sim$200 keV. The two-dimensional plots energy loss $\Delta$E vs. residual energy E$_r$ and $\Delta$E$_0$ vs. $\Delta$E are shown in Fig. \ref{fig1}a and Fig. \ref{fig1}b, respectively. The excellent energy resolutions of both $\Delta$E and E$_r$ detectors allowed unambiguous identification of $A$ and $Z$ of each product. The panel (a) is mainly related to the transfer from the projectile to the target, whereas the panel (b) is completely related to the transfer from the target to the projectile.

\begin{figure}[htp]
\begin{center}

\includegraphics[width=160mm]{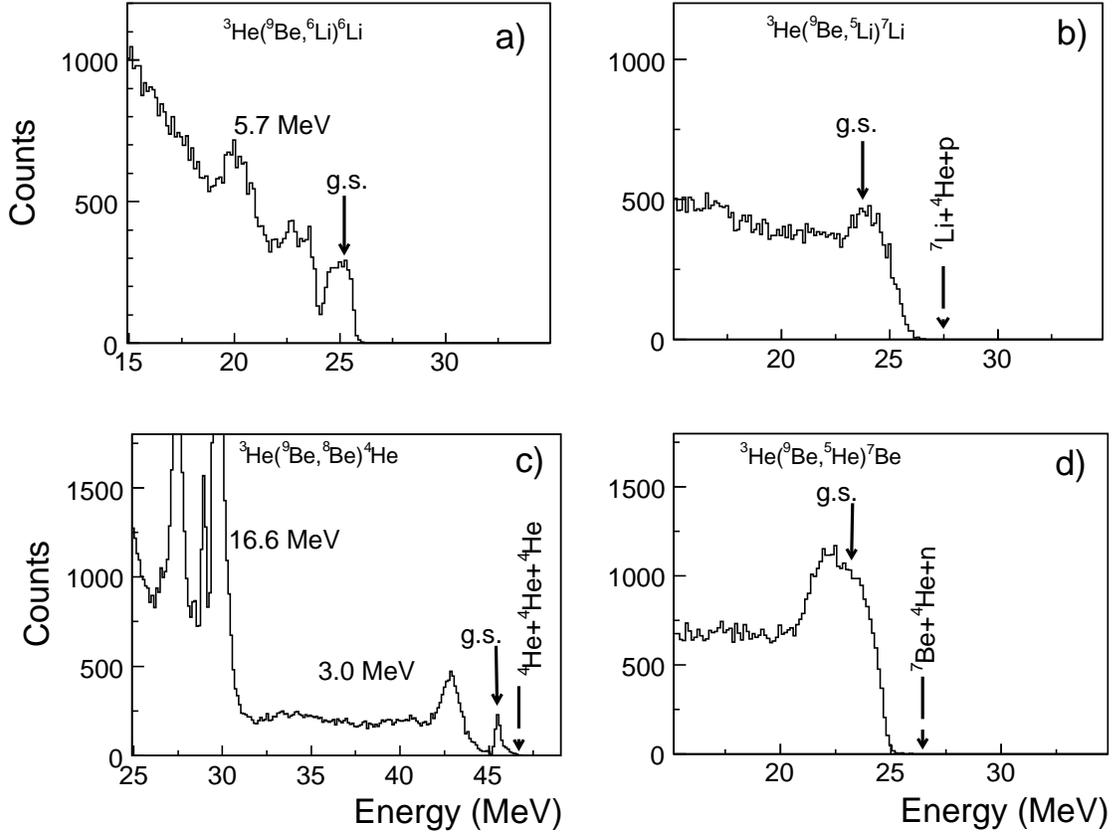}
\caption{
Spectra of the total deposited energy for the detected $^6$Li (a), $^7$Li (b), $^4$He (c) and $^7$Be (d) measured at 18$^\circ$ in the laboratory system for the reaction $^3$He(30 MeV)+$^9$Be.
}\label{fig2}

\end{center}
\end{figure}

Fig. 2 shows experimental spectra of the total deposited energy for the detected $^6$Li, $^7$Li, $^4$He and $^7$Be measured at 18$^\circ$ in the laboratory system for the reaction $^3$He(30 MeV)+$^9$Be. The plotted total energies were calculated as the sum of the calibrated energy losses $\Delta$Es and the residual energy E$_r$. The ground and the most populated excited states of $^6$Li and $^8$Be as well as the ground states populated in the reaction channels $^9$Be($^3$He,$^5$Li)$^7$Li and $^9$Be($^3$He,$^5$He)$^7$Be were unambiguously identified.

The population of the excited states of $^5$He was studied in the reaction channel $^9$Be($^3$He,$^5$He)$^7$Be together with the population of the excited states of $^6$Li in the reaction channel $^9$Be($^3$He,$^6$Li)$^6$Li as a reference case. For the case of $^6$Li one may see the known levels of this nucleus, and the peak width has value $\sigma$=200 keV equal to the detector resolution.

Since $^5$He, $^5$Li and $^8$Be are unbound nuclei and decay into $^4$He+$n$, $^4$He+$p$ and $^4$He+$^4$He, respectively, we calculated the positions in the measured spectra (Fig. 2) of the maximum energies of $^7$Li, $^4$He and $^7$Be assuming either two bodies or three bodies in the exit channels. Calculations were performed with the aid of the NRV server \cite{NRV}. The solid arrows show the maximum  values of the kinetic energies of the detected $^6$Li, $^7$Li, $^4$He and $^7$Be which correspond in the case of $^7$Li, $^4$He and $^7$Be to the two-body character of the reaction in the exit channels: $^9$Be($^3$He,$^5$Li)$^7$Li, $^9$Be($^3$He,$^8$Be)$^4$He and $^9$Be($^3$He,$^5$He)$^7$Be, respectively. The maximum values of the kinetic energies in the case of three-body final states in the reaction channels $^9$Be($^3$He,$^4$He+$p$)$^7$Li, $^9$Be($^3$He,$^4$He+$^4$He)$^4$He and $^9$Be($^3$He,$^4$He+$n$)$^7$Be are shown by the broken arrows. As can be seen from Fig. 2, the local maxima at the g.s. are close to the values corresponding to the two-body exit channels in all studied reaction channels. This indicates that the two-body reaction channels dominate leading to the unbound $^5$He, $^5$Li and $^8$Be in their ground states. Therefore, it may certainly be concluded that one-step multi-nucleon transfer is the main reaction mechanism leading to the binary exit channels.

It is expected that the experimental information provided at the end of the de-excitation process shows the strongest fingerprints of the last stages, while the influence of earlier stages and the situation just after the first reaction step are strongly masked. We can even assume that in multi-step reactions there are no fingerprints of the intermediate products in the energy distribution.

All multi-nucleon transfer processes will be described as a cluster transfer. The nucleons participating in the transfer are transferred tightly bound together, in one step. The question as to whether the transfer of so many nucleons is adequately described by the cluster model of the reaction is left open.

In the spectrum, the broad ground state of $^5$He is overlapping with the ground and first-excited states of $^7$Be, which explains why the peak is rather broad (FWHM is  $\sim-$1.5 MeV). The real peak was not observed at large angles  and no states other than those discussed above were observed. In the case of $^5$Li (Fig. 2b) the ground state in not really prominent showing a larger value of FWHM than in the case of $^5$He. It might indicate that $^5$Li is a more loosely bound nucleus than $^5$He due to the Coulomb repulsion of the additional proton from the $\alpha$ core.

Detection of unbound nuclei such as $^5$He and $^5$Li provides the opportunity to estimate the time for energy equilibration which should be less than the lifetime of the shortest-lived nuclei, $i.e.$ $^5$Li or $^5$He with the lifetime of about a few units of 10$^{-22}$ s.

\begin{figure}[htp]

 \centering

 \begin{tabular}{c}

    \includegraphics[width=140mm]{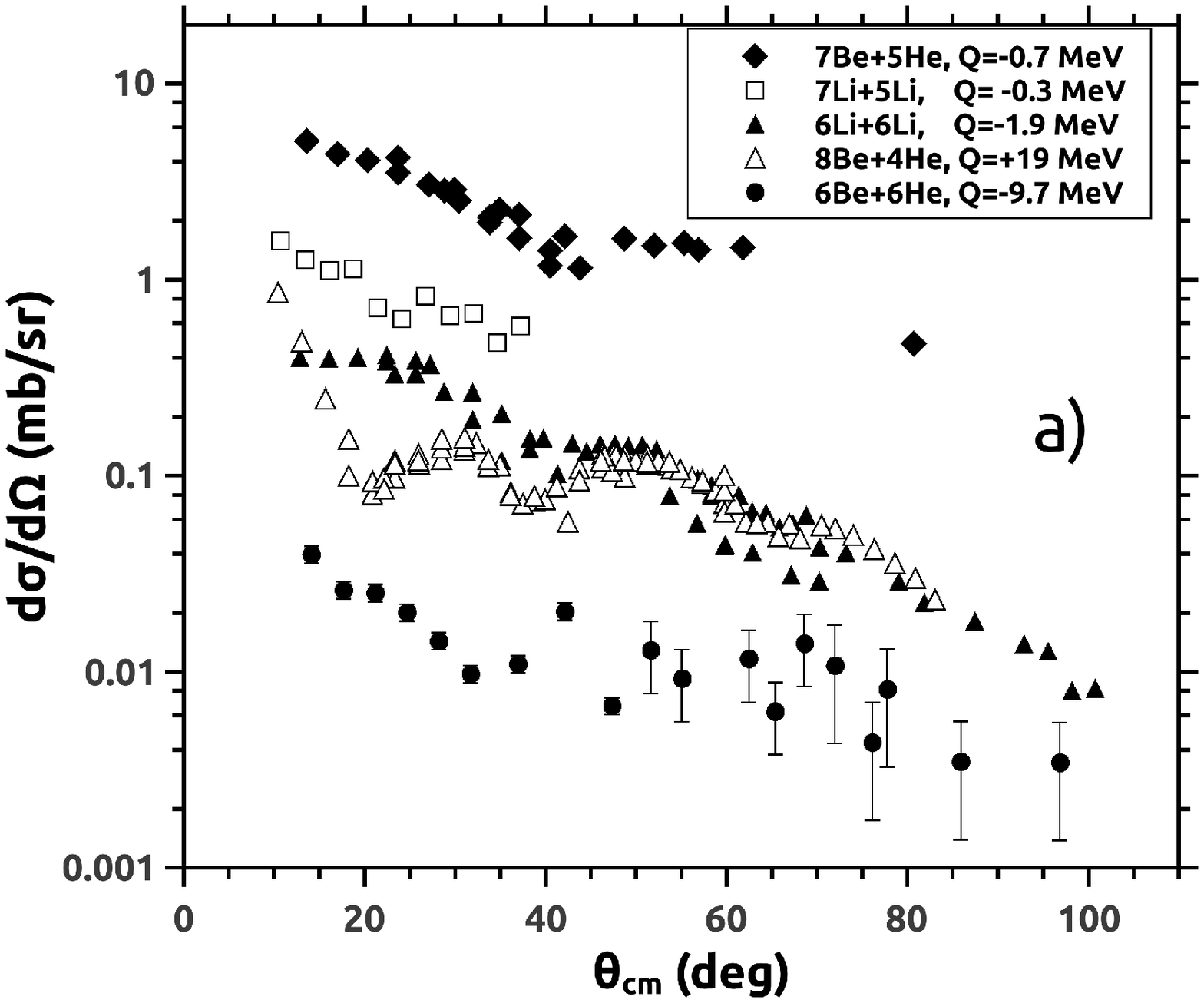}\\

    \includegraphics[width=148mm]{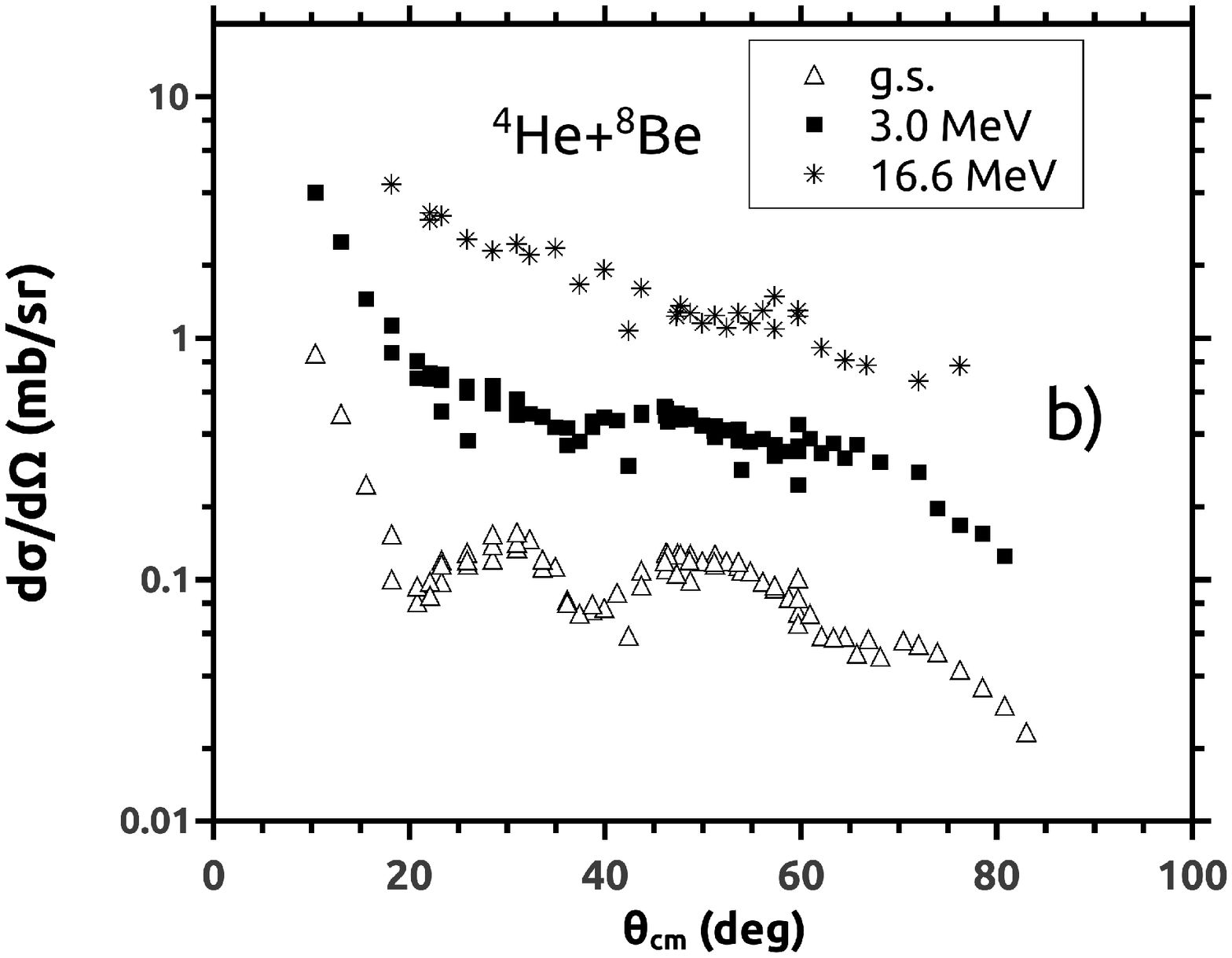}\\

  \end{tabular}

  \label{figure}\caption{
Angular distributions for 
a) 
$^3$He + $^9$Be $\rightarrow$ $^5$He + $^7$Be ($\blacklozenge$),
$^3$He + $^9$Be $\rightarrow$ $^5$Li + $^7$Li$_{g.s.}$ ($\Box$),
$^3$He + $^9$Be $\rightarrow$ $^6$Li + $^6$Li$_{g.s.}$ ($\blacktriangle$),
$^3$He + $^9$Be $\rightarrow$ $^4$He + $^8$Be$_{g.s.}$ ($\triangle$),
$^3$He + $^9$Be $\rightarrow$ $^6$He + $^6$Be$_{g.s.}$ ($\bullet$);
b) 
$^3$He + $^9$Be $\rightarrow$ $^4$He + $^8$Be$_{g.s.}$ ($\triangle$), 3.0 MeV ($\blacksquare$) and 16.6 MeV ($\ast$) excited states.
}\label{fig3}

\end{figure}

\section{Results}

\subsection{Elastic and inelastic scattering}

The	measured angular distributions for the reaction channels 
$^3$He + $^9$Be $\rightarrow$ $^5$He + $^7$Be ($\blacklozenge$),
$^3$He + $^9$Be $\rightarrow$ $^5$Li + $^7$Li$_{g.s.}$ ($\Box$),
$^3$He + $^9$Be $\rightarrow$ $^6$Li + $^6$Li$_{g.s.}$ ($\blacktriangle$),
$^3$He + $^9$Be $\rightarrow$ $^4$He + $^8$Be$_{g.s.}$ ($\triangle$),
$^3$He + $^9$Be $\rightarrow$ $^6$He + $^6$Be$_{g.s.}$ ($\bullet$) 
are shown in Fig. 3a.

First of all, one may notice that the cross section for the case of $^5$He+$^7$Be ($\blacklozenge$) is much larger than for the other reaction channels shown in Fig. 3.

The $^7$Be energy spectrum (Fig. 2d) indicates the large probability of transitions to the ground state of the $^5$He nucleus in conjunction with the 3$^-$ ground state and the $1/2^-$ first-excited (0.429 MeV) state of $^7$Be (unresolved in the present experiment). Therefore, the angular distribution for the $^5$He+$^7$Be ($\blacklozenge$) channel corresponds to the ground state and the first-excited state of $^7$Be.

 As it was mentioned, this exit channel of the reaction has the largest cross section among the channels shown in Fig. 3a.
The calculated Q-values of the investigated channels are given in the inset of Fig. 3a. It may be seen that the behaviour of the cross sections does not strictly follow the famous Q-reaction systematics. For instance, the measured differential cross section for the channel $^4$He+$^8$Be$_{g.s.}$ with the maximum value of $Q=+19$ MeV is much less than for the channel $^5$He+$^7$Be ($Q=-0.7$ MeV). The lowest cross section was observed for $^6$He+$^6$Be$_{g.s.}$ ($Q=-9.7$ MeV).

Fig. 3b shows the angular distributions of the differential cross sections for the $^3$He(30 MeV)+$^9$Be $\rightarrow$ $^4$He+$^8$Be reaction populating the ${g.s.}$, and the 3.0 MeV and around 16.6 MeV excited states. It seems that the high positive Q-value of the reaction (18.9 MeV) for this channel allows to excite very high-lying levels in the unbound $^8$Be nucleus. It is clearly visible from Fig. 3b that in the reaction $^3$He+$^9$Be $\rightarrow$ $^4$He+$^8$Be the cross section for 16.6 MeV level dominates over the other reaction channels. 

In order to describe the various reaction channels in the framework of the coupled-channel (CC) framework a special attention was paid to the elastic scattering to obtain the optical model (OM) potentials. The OM potential was chosen in the usual Woods-Saxon form
\[V(r) =  - {V_0}f(r,{R_v},{a_v}) - i{W_0}f(r,{R_w},{a_w}),\]
where the function $f(r,R,a) = {(1 + {e^{(r - R)/a}})^{ - 1}}$ and radii $R_i = r_i A^{1/3}$ depend on the mass of the heavier fragment $A$ and the corresponding reduced radius $r_i$. In addition, a spin-orbit potential with a Woods-Saxon form has been used.

In the first step, the experimental data for $^9$Be+$^3$He elastic scattering were fitted. The obtained OM potential parameters are listed in Table 1 in comparison with the parameters of Ref.~\cite{Rudchik}. The results of fitting are shown in Figs. 4 and 5 together with the experimental data.

The difference between the obtained parameters and those of Ref.~\cite{Rudchik} is probably due to the different projectile energies: our data were obtained at 30 MeV and the data of Rudchik et al. \cite{Rudchik} were obtained at 60 MeV. For instance, the depths of the real and imaginary parts are smaller at the lower incident energy, whereas the radii are larger than in Ref.~\cite{Rudchik}. 
Concerning the $^9$Be($^3$He,$^5$He)$^7$Be reaction channel, the OM parameters of both sets do not reflect any possible peculiarities due to the unbound $^5$He$_{g.s.}$, which might otherwise be expected (for example, a larger radius of this unbound nucleus considered as the candidate for one-neutron halo structure).

In the second step, of the data analysis, the cross section for the transfer channel $^9$Be($^3$He,$^4$He)$^8$Be$_{g.s.}$ was fitted and the OM potential parameters for the exit channel $^4$He+$^8$Be were obtained. In the fitting process, the OM potential parameters for the entrance channel $^9$Be+$^3$He obtained in the first step were used.

\begin{figure}[ht]
\begin{center}

\includegraphics[width=100mm]{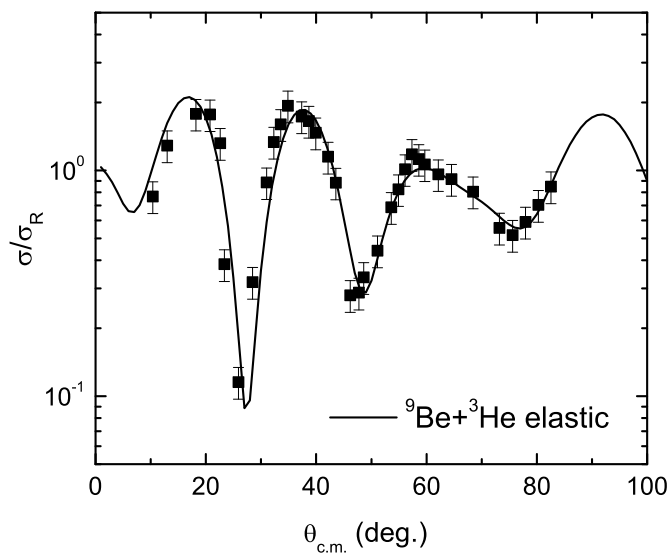}

\caption{The ratio of the measured elastic scattering cross section to the Rutherford cross section for $^3$He+$^9$Be at the incident energy 30 MeV in comparison with the OM fit.}\label{fig4}

\end{center}
\end{figure}

\begin{table}[htp]
\begin{center}
\caption{
OM potential parameters used within the optical model and CC approaches for the reaction $^3$He+$^9$Be
in comparison with the parameters of Ref.~\cite{Rudchik}. 
}

\begin{tabular}{ccccccccccc}
\br
 reaction & $V_0$ & $r_v$ & $a_v$ & $W_0$ & $r_w$ & $a_w$  & V$_{so}$ & $r_{so}$ & $a_{so}$\\
channel &  [MeV] & [fm] & [fm] & [MeV] & [fm] & [fm] &  [MeV] & [fm] & [fm]\\

\mr
$^3$He+$^9$Be & 108.5 & 1.123 & 0.54 & 15.69 & 1.15 & 0.855 & 13.63 & 1.83 & 0.4 \\
$^3$He+$^9$Be~\cite{Rudchik}& 143.4 & 1.02 & 1.10 & 38.3 & 1.405 & 1.17 &  &  &  \\
\mr
$\alpha$+$^8$Be & 90.98 & 1.382 & 0.404 & 12.36 & 1.01 & 0.4& $-$ & $-$ & $-$ \\
$\alpha$+$^8$Be~\cite{Rudchik}& 143.4 & 1.024 & 1.35 & 38.3 & 2.539 & 1.67 &  &  &  \\
$^5$He+$^7$Be  & 60.88 & 1.25 & 0.65 & 2.8 & 1.25 & 0.65 & 8.93 & 1.28 & 0.65 \\  
$^5$He+$^7$Be~\cite{Rudchik}& 122.5   & 0.9768 & 0.61 & 51.71 & 0.914 & 1.178 &  &  &  \\
$^6$Li+$^6$Li               & 110.95 & 1.307 & 0.621 & 2.48  & 1.25  & 0.65 & 1.05 & 1.25 & 0.65  \\
$^6$Li+$^6$Li~\cite{Rudchik}& 122.5 & 0.979 & 0.905 & 51.71 & 0.914 & 1.178 &  &  &  \\
\br 
\end{tabular}
\end{center}
\end{table}



\begin{figure}[htp]

  \centering

 \begin{tabular}{c}

    \includegraphics[width=90mm]{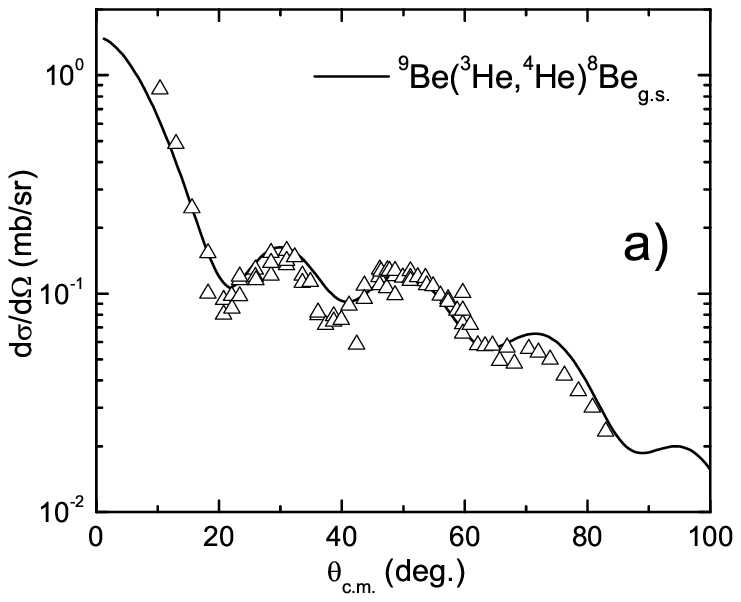} \\

    \includegraphics[width=90mm]{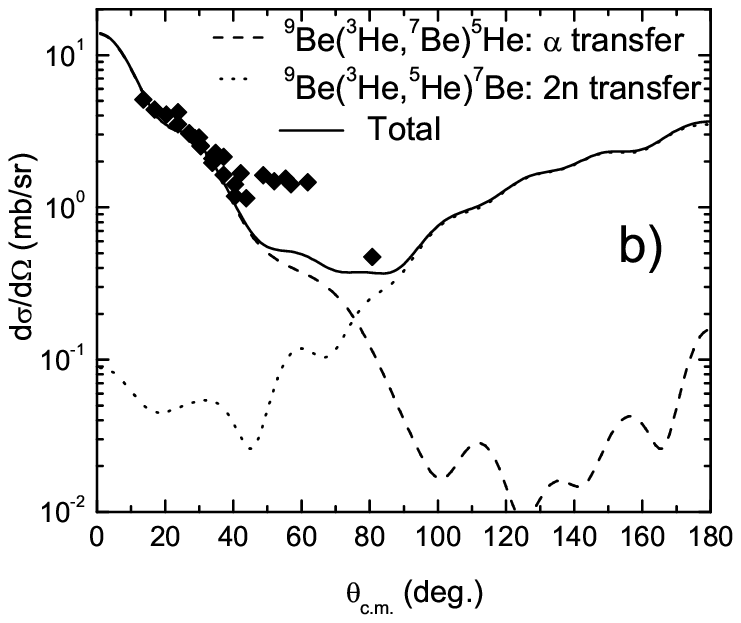} \\
\includegraphics[width=90mm]{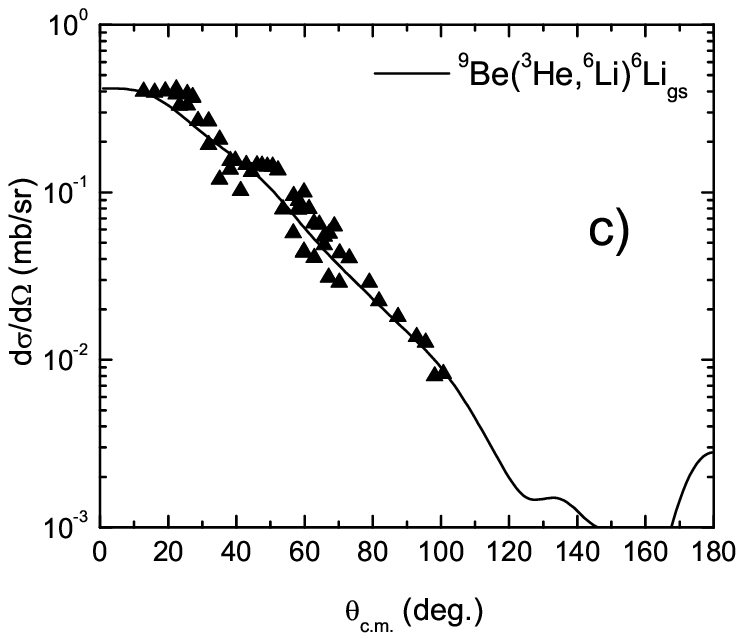} \\

  \end{tabular}

 \label{figure}\caption{
The angular distributions for a) $^3$He + $^9$Be $\rightarrow$ $^4$He + $^8$Be$_{g.s.}$ ($\triangle$), b) $^3$He + $^9$Be $\rightarrow$ $^5$He + $^7$Be ($\blacklozenge$) and c) $^3$He + $^9$Be $\rightarrow$ $^6$Li + $^6$Li$_{g.s.}$ ($\blacktriangle$). The curves are the results of the optical-model and the coupled-channel calculations (see text).
}\label{fig5}

\end{figure}

In the third step, the cross sections for the transfer reactions $^9$Be($^3$He,$^7$Be)$^5$He and $^9$Be($^3$He,$^5$He)$^7$Be were fitted simultaneously and the OM potential parameters for the exit channel $^5$He+$^7$Be were obtained (Table 1). In the fitting process, the OM potential parameters for the entrance channel $^9$Be+$^3$He obtained in the first step were used. Later in the calculations of the total cross section both transition amplitudes corresponding to the alpha-transfer mechanism (the $^9$Be($^3$He,$^7$Be)$^5$He channel) and the 2$n$-transfer mechanism (the $^9$Be($^3$He,$^5$He)$^7$Be channel) have been taken into account. All calculations and fitting have been performed using the FRESCO code~\cite{Fresco} in the framework of the CC method.
 
By fitting the experimental angular distributions, the attempt was made to determine which reaction mechanisms are the most important for the measured distributions. The results are shown in Fig. 5 for the following channels: a) $^3$He + $^9$Be $\rightarrow$ $^4$He + $^8$Be$_{g.s.}$, b) $^3$He + $^9$Be $\rightarrow$ $^5$He + $^7$Be and c) $^3$He + $^9$Be $\rightarrow$ $^6$Li + $^6$Li$_{g.s.}$.

It is clear that in the case of $^3$He+$^9$Be $\rightarrow$ $^7$Be+$^5$He reaction channel the $\alpha$-transfer (long-dashed curve) dominates at forward angles, whereas the transfer of the two neutrons (short-dashed curve) dominates at backward angles. 

The fit for $^3$He+$^9$Be $\rightarrow$ $^6$Li+$^6$Li (solid curve) was obtained assuming $t$ transfer from the target nucleus to the projectile. A good agreement between the experimental data and the calculations is observed. A similar situation may be expected for the case of $^3$He+$^9$Be $\rightarrow$ $^5$Li+$^7$Li where $d$ transfer dominates.

As could be seen from Fig. 5, the main processes describing the angular distributions of the exit channels $^4$He+$^8$Be$_{g.s.}$, $^5$He+$^7$Be and $^6$Li+$^6$Li$_{g.s.}$ are, respectively, the neutron, $\alpha$ and $t$ transfer from the target nucleus to the projectile.

\subsection{Cluster transfer in $^3$He+$^9$Be reaction}

Due to the Borromean structure of $^9$Be the breakup of this nucleus may, in principle, proceed directly into two $\alpha$ particles plus a neutron or via one of the unstable intermediate nuclei: $^8$Be or $^5$He. The attempt was made to get some insights into $n$+$^8$Be, $^4$He+$^5$He, $d$+$^7$Li and $t$+$^6$Li cluster configurations inside of $^9$Be and especially to estimate the relative strengths of the most interesting $^8$Be+$n$ and $^5$He+$\alpha$ cluster configurations in $^9$Be.

Most likely the studied reaction channels result from the cluster transfer:
\begin{description}
\item{ $^3$He + $^9$Be $\rightarrow$ $^3$He + ($n$ + $^8$Be) $\rightarrow$ ($^3$He + $n$)+ $^8$Be $\rightarrow$  $^4$He + $^8$Be
\item $^3$He + $^9$Be $\rightarrow$ $^3$He + ($\alpha$ + $^5$He) $\rightarrow$ ($^3$He + $\alpha$) + $^5$He $\rightarrow$  $^7$Be + $^5$He 
\item $^3$He + $^9$Be $\rightarrow$ $^3$He + ($t$ + $^6$Li) $\rightarrow$ ($^3$He + $t$) + $^6$Li $\rightarrow$  $^6$Li + $^6$Li 
\item $^3$He + $^9$Be $\rightarrow$ $^3$He + ($d$ + $^7$Li) $\rightarrow$ ($^3$He + $d$) + $^7$Li $\rightarrow$  $^5$Li + $^7$Li 
\item $^3$He + $^9$Be $\rightarrow$ $^3$He + ($^3$He + $^6$He) $\rightarrow$  ($^3$He + $^3$He) + $^6$He $\rightarrow$  $^6$Be + $^6$He} 
\end{description}

These reaction channels are of specific interest since they provide direct information on the cluster structure of the initial and final states.

Based on the angular distributions shown in Fig. 3, one may estimate the probability of cluster configurations in $^9$Be. To obtain the branching ratios (BRs) for the observed channels, the angular distributions were integrated over the solid angle. The probability was calculated as weighted integrals among the channels listed above. The obtained values of the BRs are given in Table 2.

\begin{table}[h]
\centering
\caption{Probability of cluster configurations in the $^9$Be nucleus}
\footnotesize\rm
\begin{tabular}{lll}
\br
Exit channel & $^9$Be cluster configuration &  Branching ratio (\%) \\
\mr
$^4$He + $^8$Be  & $n$ + $^8$Be & $\le$68.7 $\pm 10$ \\
$^5$He + $^7$Be  & $\alpha$ + $^5$He & $\le$25.1 $\pm 5$ \\
$^6$Li + $^6$Li$_{g.s.}$ & $t$ + $^6$Li & $\ge$3.3 $\pm 2$   \\
$^5$Li + $^7$Li$_{g.s.}$ & $d$ + $^7$Li & $\ge$2.7 $\pm 2$  \\
$^6$Be + $^6$He$_{g.s.}$ & $^3$He + $^6$He & $\ge$0.2 $\pm 0.7$  \\
\br 
\end{tabular}
\end{table}

The ratio of $^8$Be+$n$ to $^5$He+$\alpha$ is 2.7, which is close to the ratio 2:1 given in Ref.~\cite{Nyman} and contradicts the conclusion of Ref.~\cite{Grigorenko} where it was reported that the contribution of the $^5$He+$\alpha$ configuration is negligible.

The obtained result is partially consistent with the results of Ref.~\cite{Charity} confirming that decay of ${}^9$B has the dominant branch $\alpha$+${}^5$Li implying that “the corresponding mirror state in ${}^{9}$Be would be expected to decay through the mirror channel $\alpha$+${}^{5}$He, instead of through the $n$+${}^{8}$Be(2$^+$) channel”.

As for the conclusion of the more detailed experiment Refs.~\cite{Brown, Papka} where measurements showed that the $^5$He+$\alpha$ channel was not important for the low-lying excited states of $^9$Be, the discrepancy may be explained from the experimental point of view. Our data are based on inclusive measurements, whereas the experimental results of Refs.~\cite{Brown, Papka} are based on exclusive measurements. The latter always requires fully defined kinematics allowing the complete reconstruction and identification of events. The values of the BR given in Ref.~\cite{Brown} were corrected by the geometrical detector efficiency obtained using a special simulating code.

Additionally, it should be noted that the values of the branching ratios given in Ref.~\cite{Brown} are assigned especially to the excited states, whereas our values of the branching ratios are associated entirely with the overall nuclear system $^9$Be. This clearly calls for theoretical calculations of the $^8$Be+$n$ and $\alpha$+$^4$He reaction rate, which at the moment represents a great theoretical challenge.

The direct triton pick-up process requires that $^9$Be has the ($^6$Li+$t$) cluster configuration, which seems highly unlikely compared to the ($\alpha$+$\alpha$+$n$) configuration. Also, the interpretation of any ($^3$He,$^6$Li) reaction in terms of triton pick-up requires that $^6$Li has the ($^3$He +$t$) cluster configuration, which seems equally unlikely compared to the ($\alpha$ + $d$) configuration. It has been shown in Ref.~\cite{Young} that for $^6$Li the ($^3$He+$t$) configuration is only slightly less probable than the ($\alpha$+$d$) configuration. Since such cluster configurations are not mutually exclusive, one cannot judge upon the relative importance of the ($^6$Li+$t$) cluster configuration of $^9$Be without making careful cluster structure calculations. Indeed, the differential cross section for the $^7$Li in the outgoing channel is higher than for the channel leading to $^6$Li.

Another objective in measuring the ($^3$He,$^6$Li) channel was to evaluate the strength of the triton clustering. The present measurements imply that $^9$Be contains a significant fraction of the ($^6$Li+$t$) cluster configuration. The same is true for the $^3$He-clustering related to the reaction channel $^3$He+$^9$Be$\rightarrow$$^6$He+$^6$Be, i.e. $^6$He+$^3$He clustering in $^9$Be. Similarly, the $^3$He+$^3$He cluster structure could be assigned to $^6$Be, $t$+$^3$He to $^6$Li, $^3$He+$^4$He to $^7$Be as well as $^5$He+$^4$He and $^3$He+$^6$He to $^9$Be.

As estimated above, the time for the energy equilibration is rather short. Therefore, it could be assumed that the sequential multi-step reactions cannot not take place and all transfers must happen rather suddenly.

\section{Conclusions}
The angular distributions of the $^9$Be($^3$He,$^3$He)$^9$Be, $^9$Be($^3$He,$^5$He)$^{7}$Be, $^9$Be($^3$He,$^5$Li)$^7$Li, $^9$Be($^3$He,$^6$Be)$^6$He and $^9$Be($^3$He,$^6$Li)$^{6}$Li reaction channels were measured and described within the framework of the optical model, the coupled-channel approach and the distorted-wave Born approximation.

The performed analysis of the experimental data shows that the potential parameters are quite sensitive to the exit channel and hence to the cluster structure of the populated states, which allows to make general observations and conclusions regarding the internal structure of the target and residual nuclei.

In the first step, the experimental data for $^9$Be+$^3$He elastic scattering were fitted. The fitting parameters were obtained and were compared with the previous measurements. In the second step, the cross sections for the transfer reactions $^9$Be($^3$He,$^4$He)$^8$Be$_{g.s.}$ and $^9$Be($^3$He,$^5$He)$^7$Be were fitted. The fitting result does not reflect any possible peculiarities in the OM parameters in the case of the unbound $^5$He nucleus.

The cross sections for $n$, $d$, $t$, and $^3$He transfer channels were measured in the $^3$He+$^9$Be collision at the incident energy 30 MeV. The proposed OM potential provides a good fit of the elastic scattering in the entrance channels. The DWBA calculations agree well with the transfer reaction data. The spectroscopic factors needed to fit the transfer channels are close to unity, which confirms the significant contribution of the considered cluster configurations to the structure of the ground states. We also show the possibility to extract the structural information from the comparative analysis of the $^9$Be($^3$He,$^6$Li)$^6$Li and $^9$Be($^3$He,$^6$Be)$^6$He reaction channels.

The experiment was designed to study the breakup of $^9$Be in an attempt to determine the contribution of the $^8$Be+$n$ and $^5$He+$\alpha$ channels in the inclusive measurements. The ratio 2.7:1 may be assigned for the contributions of these two channels, respectively. The determined branching ratios confirm that the $^5$He+$\alpha$ breakup channel plays a significant role. The inclusive experimental method used has the advantage of its experimental simplicity in comparison with the exclusive one ~\cite{Brown, Papka}.

\section*{Acknowledgments}
We would like to thank the JYFL Accelerator Laboratory and NPI (\v{R}e\v{z}) for giving us the opportunity to perform this study as well as the cyclotron staff of both institutes for the excellent beam quality. This work was supported in part by the Russian Foundation for Basic Research (project numbers: 13-02-00533 and 14-02-91053), the CANAM (IPN ASCR) and by the grants to JINR (Dubna) from the Czech Republic, the Republic of Poland and the mobility grant from the Academy of Finland.

\section*{References}


\end{document}